\documentclass[12pt]{article}\usepackage[hyperfootnotes=false]{hyperref}
\usepackage{epsfig}
\usepackage{amsmath}
\usepackage{amssymb}
\usepackage{graphicx}
\newlength{\abstractwidth}
\renewcommand{\thefootnote}{\fnsymbol{footnote}}
\renewcommand{\thanks}[1]{\footnote{#1}} % Use this for footnotes
\newcommand{\starttext}{
\setcounter{footnote}{0}
\renewcommand{\thefootnote}{\arabic{footnote}}}
\renewcommand{\theequation}{\thesection.\arabic{equation}}
\newcommand{\be}{\begin{equation}}
\newcommand{\bea}{\begin{eqnarray}}
\newcommand{\eea}{\end{eqnarray}}
\newcommand{\beq}{\begin{equation}}
\newcommand{\ee}{\end{equation}}
\newcommand{\eeq}{\end{equation}}

\def\ba{\begin{eqnarray}}
\def\ea{\end{eqnarray}}

\def\12{{1 \over 2}}
\def\eq{&=&}

\def\ra{\rangle}

\def\simleq{\; \raise0.3ex\hbox{$<$\kern-0.75em
\raise-1.1ex\hbox{$\sim$}}\; }
\def\simgeq{\; \raise0.3ex\hbox{$>$\kern-0.75em
\raise-1.1ex\hbox{$\sim$}}\; }

\def\O2{\Omega_2}

\def\bi{\begin{itemize}}
\def\ei{\end{itemize}}

\def\sc{\setcounter{equation}{0}}

\def\W{$\Omega$}
\def\W'{$\Omega$}

\def\V{\Omega}
\def\V'{\Omega}

\def\c{{\cal{C}}}

\def\bn{\bigskip \noindent}

\begin{document}
\renewcommand{\theequation}{\thesection.\arabic{equation}}

\begin{titlepage}
\rightline{}
\bigskip
\bigskip\bigskip\bigskip\bigskip
\bigskip
\centerline{\Large \bf { Addendum to }}
\bigskip

\centerline{\Large \bf { Computational Complexity and Black Hole Horizons }}

\bigskip

\begin{center}
\bf Leonard Susskind \rm

\bigskip

Stanford Institute for Theoretical Physics and Department of Physics, \\
Stanford University,
Stanford, CA 94305-4060, USA \\
\bigskip

\bigskip

\end{center}
\bigskip\bigskip
\bigskip\bigskip
\begin{abstract}

In this addendum to [arXiv:1402.5674] two points are discussed. In the first additional evidence is provided for a dual connection between the geometric length of an Einstein-Rosen bridge and the computational complexity of the quantum state of the dual CFT's.
The relation between growth of complexity and Page's ``Extreme Cosmic Censorship" principle is also remarked on.

The second point involves a gedanken experiment in which Alice measures a complete set of commuting observables at her end of an Einstein-Rosen bridge is discussed. An apparent paradox is resolved by appealing to the properties of GHZ tripartite entanglement.

\medskip
\noindent
\end{abstract}
\end{titlepage}

\starttext \baselineskip=17.63pt \setcounter{footnote}{0}

\tableofcontents

\sc
\section{Complexity and Wormhole Length}

\subsection{Evidence from Shockwave Geometries}

In \cite{Susskind:2014rva}  a relation between computational complexity and the lengths of  Einstein-Rosen bridges (ERB) was conjectured.

 In the context of the two-sided eternal black hole, define the quantum state 
  
  $$|\Psi (t_L, t_R)\ra$$
  
 \bn
 to be the state of the boundary system at left and right times $t_L, t_R.$ Let $\c(t_L,t_R)$ be the computational complexity of the state.
  The conjecture stated that in ADS units, the length $d$ of the bridge  between two boundary points labeled $t_L$ and $t_R$ is proportional  to ratio of  (computational complexity) to (entanglement entropy) of the state  $|\Psi (t_L, t_R)\ra$,
  
  \be
  d(t_L, t_R) = \c(t_L,t_R)/S
  \ee

\bn
The evidence for the relation was that for the thermofield-double  (TFD ) state both quantities grow linearly with time until the classical recurrence time. In the first part of this paper I will provide additional evidence that the length-complexity relation continues to be correct for Shenker-Stanford shockwave geometries \cite{Shenker:2013pqa}. The analysis will be for  the $(2+1)$-dimensional BTZ case. Throughout the paper I will use the conventions and notations of \cite{Susskind:2014rva}.

Let's review the TFD case. The TFD state is simple in the sense that it has very little ``vertical" entanglement. A rough description of it is as a system of $S$ entangled Bell pairs shared between Alice (on the left side) and Bob (on the right side). There is maximal ``horizontal" entanglement between Bob's share and Alice's share. But the Bell pairs are not vertically entangled with each other.

However, as time evolves the state becomes more complicated. If we partition the qubits by grouping half of Alice's with half of Bob's, then after a time this half will become vertically entangled with the remaining half.
The increase of vertical entanglement is a manifestation of increasing complexity.

The vertical entanglement can be evaluated using the Ryu-Takayanagi \cite{Ryu:2006bv} minimal surface hypotheses as explained by Hartman and Maldacena \cite{Hartman:2013qma}. One
begins by dividing the  boundary  $(D-2)$-spheres  into hemispheres. Next a  $(D-2)$  dimensional minimal surface
is constructed that goes through the bulk and terminates on the two equators at the boundaries. The vertical entanglement is the area of that surface.

Initially the minimal surface is a single connected component connecting the left and right side, but at a time of order $l_{ads}$ a transition occurs, and the surface ``snaps" into two disconnected components. At that time the leading $N$ dependence of the vertical entanglement becomes maximal and stops increasing. Therefore it can no longer be used as a measure of increasing complexity, or of distance across the ERB.

However, although  the connected surface ceases to be the absolute extremal surface, the connected surface does not disappear.  In \cite{Susskind:2014rva} it was interpreted as representing the computational complexity of the state, which can continue to increase far beyond the maximal vertical entanglement. We may also use it to represent the distance across the ERB.

In this addendum I will  consider the $(2+1)$-dimensional case where the relevant formulas have been worked out in \cite{Shenker:2013pqa}. The minimal surfaces become minimal lines and the Hartman-Maldacena-Ryu-Takayanagi (HMRT) surfaces are space-like geodesics.

 There is one point that requires a brief comment. The distance between the boundaries diverges because the metric blows up at the boundaries. This divergence is fully understood and should be regulated. If we define points at radial coordinate $r$ instead of at the actual boundary, then the geodesic distance in the TFD case is given  in terms of $R,$ the radius of the horizon; $r,$ the cutoff distance; and the times $t_L, \ t_R,$ \cite{Shenker:2013pqa}\footnote{There is a discrepancy between the notations of \cite{Susskind:2014rva} and \cite{Shenker:2013pqa} with respect to the sign of $t_L.$ In this paper I will use the convention of \cite{Susskind:2014rva} in which all times increase toward the future. To compare with \cite{Shenker:2013pqa} the sign of $t_L$ must be changed.}

 \be
 \frac{d}{l_{ads}} = 2 \log{\frac{2r}{R}} + 2\log{\cosh{\frac{R}{2l_{ads}^2}(t_R + t_L)}}
 \ee

\bn
 The infinity is isolated in the first term when $r\to \infty.$ It is universal and may be dropped.

 The second term shows that the bridge-length quickly becomes a linear function of time.

 From now on I will follow \cite{Susskind:2014rva}  and take the radius $R$ to be equal to the ADS scale,

 $$R=l_{ads}$$

 \bn
 Then dropping the infinite term the length is given by
  \bea
 \frac{d}{l_{ads}} \eq 2\log{\cosh{\frac{(t_R + t_L)}{2l_{ads}}}} \cr \cr
 &\to &  \frac{t_R+t_L}{l_{ads}}
 \label{length}
 \eea

\bn
 If we assume the two sides evolve as independent quantum circuits and that the computation rate is given by equation 3.5 of \cite{Susskind:2014rva} then \ref{length} also represents the growing computational complexity of the state  $|\Psi (t_L, t_R)\ra.$

 The growth of complexity cannot proceed beyond the maximal complexity $\c_{max} = e^S.$ There is independent evidence that the classical geometry of the ERB must break down at such large time.

 The evidence for a connection between complexity and ERB length is not very persuasive since it only applies to the TFD state. One would clearly like to test it  in a wider context. Toward that end we consider Shenker-Stanford shockwave geometries. Let's begin with the computational complexity side of things.

 First consider the computational complexity of the TFD state itself. Insofar as it may be modeled by $S$ Bell pairs, its complexity is of order  $S.$ The bridge length is identified with the complexity divided by the entropy and is therefore of order $1.$  The TFD corresponds to the state $|\Psi (t_L, t_R)\ra$ at $t_L = t_R =0.$ From \ref{length} we see that the length of the bridge is also order $1.$

 Now suppose Alice acts with a precursor to  create a shockwave at fictitious time $t_L = -t_W$ (In Shenker-Stanford's notation this would be $t_L = +t_W$). This adds a complexity to the state given by the number of gates needed to implement the precursor operator

 \be
 W_p = U(t_W) W U^{\dag}(t_W).
 \label{precursor}
 \ee

\bn
 In \cite{Susskind:2014rva} I explained that the complexity of the precursor was of order the complexity of $U,$ but now I want to be more precise. The argument illustrated in figure 1 of \cite{Susskind:2014rva} shows that for large $t_W$ the complexity of \ref{precursor} is \it twice  \rm the complexity of $U$ because of the presence of both  $U$ and $U^{\dag}.$ Moreover, it was argued that as a consequence of the chaotic nature of the system, there is no cancelation between the two.
 Thus the complexity due to \ref{precursor} is

 \be
 \c_p = 2 S \frac{t_W}{l_{ads}}
 \ee

\bn
 The factor of $2$ was not important in estimating the complexity of a precursor, but it is critical for comparing the complexity of the state with the bridge length. To do that we go back to equation 19 of reference \cite{Shenker:2013pqa}, which  is accurate for large $t_W.$  I will re-write the equation in the notation of this paper:

\be
\frac{d}{l_{ads}} = 2 \log{   \left[      \cosh{\frac{t_R+t_L}{2l_{ads}}}    + q e^{(2t_W  +t_L - t_R )/2l_{ads}  }                   \right]    }
 \ee

\bn
 where $q$ is a small number of order the ratio of the temperature and the mass of the black hole.

 Using the symmetry of the TFD state under translations generated by $(H_R - H_L),$ with no loss of generality we may set $t_R=0.$

 \be
 \frac{d}{l_{ads}} = 2 \log{\left[  \cosh{\frac{t_L}{2l_{ads}}}    + q e^{(2t_W  +t_L  )/2l_{ads} }
  \right]}
\label{master}
\ee

\bn
For large $t_W$ \ref{master} becomes

 \be
d =  2t_W  +t_L
\label{baster}
\ee

\bn
This is precisely what one expects for the complexity. The $2t_W$ is the term from the precursor; the $2$ representing the fact that the precursor contains both  $U$ and $U^{\dag}.$ The dependence on $t_L$ is the subsequent growth of complexity as $t_L$ increases.

This  result can be generalized in several directions. One is to consider shockwaves from the left and the right simultaneously. Both Bob and Alice activate precursors which create shockwaves originating from times $-t_{W_L}$ and $-t_{W_R}.$ The complexity-to-entropy ratio of the state $$|\Psi (t_L, t_R)\ra$$ in this case is expected to be

\be
\c/S =( 2t_{W_L} + 2t_{W_L}) + (t_L +t_R).
\ee

\bn
The first term comes from the contribution of the precursors and the second, from the later growth of complexity. An analysis of the two-shockwave geometry  \cite{Shenker:2013yza} gives the same formula for the length of the geodesic connecting the boundary points\footnote{Douglas Stanford, Private communication.}.

These calculations are obviously not conclusive but they do support the identification of complexity and ERB-length. Additional data can be obtained by considering more general shockwave geometries, as well as higher dimensional examples. One warning about the latter case: the length should be defined by the area of the appropriate HMRT surfaces, not the length of  geodesics.

\subsection{Stretching and Extreme Cosmic Censorship}

The motivation for suggesting a relationship between ERP growth and computational complexity was to establish a dual description of what I called  "the stretching hypothesis"  \cite{Susskind:2014rva}\footnote{The importance of stretching as a protection against firewalls was emphasized by J. Maldacena.} The stretching hypothesis states that black holes are formed in such a way that the interior geometry grows rather than shrinks. Both are mathematically possible, but according to the hypothesis, black holes formed by natural processes are of the growing kind.

Relating the size of the interior geometry to complexity allows a restatement of the hypothesis:

\bn
\it Black holes are formed in such a way that the complexity of the state  increases. \rm

\bn
Since, by time reversal, there are just as many states in which complexity is decreasing, as there are states with
increasing complexity, the hypothesis throws away half the states in the Hilbert space\footnote{For the closed ADS system
  these discarded states will recur  on quantum-recurrence time-scales. }.

The stretching hypothesis is probably very closely related to Page's recent ``Extreme Cosmic Censorship" conjecture \cite{Page:2013mqa} that excludes states which trace back to past singularities. Again this throws away half the initial states. The two ideas can be related through the following observation. Past singularities, being time reversals of future singularities, are states of extremely high computational complexity, but in which the complexity  decreases with time.

The main difference between \cite{Page:2013mqa} and \cite{Susskind:2014rva}, is that \cite{Susskind:2014rva} treats the increase of complexity, not as an absolute law, but as a ``quasi-law" in the same sense that the second law of thermodynamics is quasi; entropy \it almost \rm always increases. But local violations can take place and with enough effort, reversals of the second law can be made to happen in a local region of space. That's what happens when Alice acts with a highly complex precursor.

\sc
\section{Measurements and GHZ States}

On an entirely different matter, I want to discuss a puzzling point which has come up in several conversations.
Suppose that Alice uses her powers to measure a complete set of commuting operators describing the left-side black hole. There are two questions. The first is whether this creates a firewall on Bob's side. The second, assuming that the answer to the first question is no, is whether Alice can subsequently send a signal to Bob.

To set up the problem we must expand the system to include Alice's laboratory. We may think of the CFT's as describing the zones of the two black holes, and the laboratory as being the regions beyond the zones. Alice lives in such an outer region but can interact with the zone and stretched horizon of her black hole. Let us suppose that Alice's quantum computer has a memory consisting of a set of $S$ qubits. A measurement is an interaction between the memory-qubits and the qubits describing the left-side black hole. We may assume that each memory-qubit interacts, by means of a single gate,  with one black hole qubit. If the black hole qubit is in the state $|0\ra$ the corresponding memory-qubit will record $|0\ra$ and similarly if the black hole qubit is $|1\ra$ the memory-qubit will record $|1\ra.$

We will imagine that Alice has access to the information in the memory. We may even assume that the memory is in Alice's brain.
But we will also consider another participant, Charlie, who for the most part, is passive. From Charlie's point of view the system consists of three parts:

\bi
\item The right-side black hole $R.$

\item The left-side black hole $L$

\item Alice and her memory $M.$

\item The overall left side, consisting of union of  $L$ and $M,$  is called $LM.$

\ei

Let's answer the first question; does Alice's measurement create a firewall at Bob's end. The measurement can be described as a unitary operator acting on the $LM$ system.
The number of gates required to carry out the measurement of all $S$ qubits in Alice's black hole is $S.$ Thus the measurement introduces a complexity no larger than $S.$ As explained in \cite{Susskind:2014rva}, this does not reach deeply enough into the near-horizon region to send a signal to Bob. To send a shockwave the minimum complexity is the scrambling value, $S \log S.$

Let us consider the second question. Once the measurement has been made, the system becomes an entangled tripartite system consisting of $R,$ $L,$ and $M, $ Alice herself being thought of as part of the memory system. The question is:
 Now that Alice has made a complete measurement, can she, by manipulating herself,  her memory, and possibly the left black hole, send a signal to Bob?
 According to her reckoning, by doing the measurement she has collapsed the wave function, so that  the left and right sides are no longer entangled. Thus she must conclude that she cannot send a signal.

 But let's view it from Charlie's perspective. Charlie says that Alice's actions were local operations on the overall left side, and could not possibly have changed the entanglement between $R$ and the union $LM.$ Therefore $LM$ must still be connected by an ERB to $R.$  The ERB connects all three systems---right black hole; left black hole; and memory.  This, in principle should allow a signal to be sent to Bob. This seems to contradict Alice's analysis.

 To analyze the system from Charlie's point of view we have to introduce a factor to the Hilbert space to describe $M.$  This factor must be large enough to store $S$ bits. We may represent it by another set of $S$ qubits.

 Let's begin with the TFD state which we model by a maximally entangled state of the left and right black holes. The tensor factor representing the memory begins in the state $|000000000....\ra.$ Thus the initial state is,

 \be
 |in\ra = \left[ |00\ra + |11\ra \right ]^{\otimes S}  \otimes |000000000....\ra
 \ee

 \bn
 where the last factor represents the memory before the measurement.

 When the measurement takes place the memory-qubits become correlated with the left qubits, but it is easy to see that the final state is symmetric between all three systems. It has the form of a tensor product of $S$ GHZ-triplets. Define a GHZ triplet to have the form,

 \be
 |ghz\ra = |000\ra + |111\ra.
 \ee

 \bn
 After the measurement the state is

 \be
 |GHZ\ra =  |ghz\ra ^{\otimes S}.
 \ee

\bn
 GHZ states have a particular form of tripartite entanglement, for which the ERB connecting the three parties cannot be described by conventional classical geometry.

 Here are the relevant properties of such states.

 \bn
 1) Any one of the parties in a GHZ state is maximally entangled with the union of the other two. Thus Bob's black hole is maximally entangled with the union of the memory and the left black hole. Similarly the memory is maximally entangled with the union of left and right black holes.

 \bn
2)  When any one of the three parties is traced over, the remaining bipartite density matrix is separable. This means that there is no bipartite entanglement between pairs of parties.

\bn

\begin{figure}[h!]
\begin{center}
\includegraphics[scale=.5]{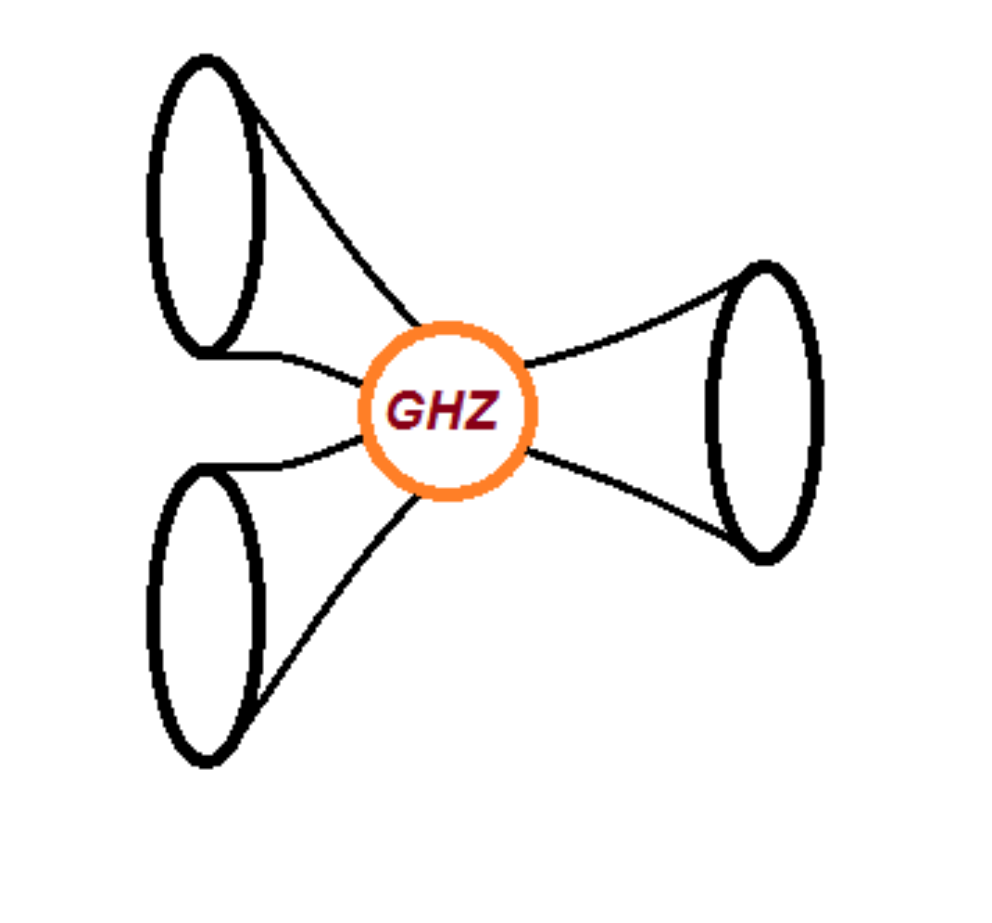}
\caption{ERB for a GHZ-entangled system of three black holes. The core of the ERB must have properties that cannot be described
by classical geometry. }
\label{ghz}
\end{center}
\end{figure}

How these two properties can be described by some kind of generalized ERB is unknown but I will assume that it makes sense to do so (see figure \ref{ghz}).

Now let us consider what Alice can and cannot do, as far as signaling Bob is concerned. She clearly cannot send a signal by manipulating the memory (which can mean herself as well as any other hardware) since she is completely unentangled with Bob's black hole.

Here is something else she cannot do. She can't take her record around to Bob's side and jump in with the record, without disturbing Bob's black hole massively. The reason is that the record has $S$ bits of information and simply cannot fit in the original black hole. In order to fit in the black hole Alice would have to leave behind almost all knowledge of the experimental outcome.

However, let us consider what Charlie can do, assuming he has a powerful quantum computer. We know that Bob's black hole is maximally entangled with the union of left black hole, and the memory, i.e., the $LM$ system. Therefore, he should be able to do a complex process on the $LM$ system that can send a signal to Bob.

To do that, Charlie must activate a very complex precursor. In fact the precursor must involve both $L$ and $M$ is a way that \it undoes \rm the measurement, and erases Alice's memory. In normal circumstances we consider measurements to be irreversible because of their complexity. But in principle anything that happens can be made to un-happen \cite{Bousso:2011up}.

Therefore there is no contradiction between Alice's description and Charlie's. Alice correctly believes that she cannot send a signal from her subsystem after the measurement. But Charlie believes that he can send Bob a signal by applying a precursor complex enough to erase the memory of the experiment.

\section*{Acknowledgements}

I am grateful to Douglas Stanford for help understanding shockwave geometries and the length of the geodesics through them.

I also thank Raphael Bousso for a discussion about the interpretation of Alice's measurement.

Support for this research came through NSF grant Phy-1316699 and the Stanford Institute for Theoretical Physics.

\end{document}